\documentclass[conference]{IEEEtran}
\IEEEoverridecommandlockouts
\usepackage{cite}
\usepackage{amsmath,amssymb,amsfonts}
\usepackage{algorithmic}
\usepackage{graphicx}
\usepackage{textcomp}
\usepackage{xcolor}
\usepackage{booktabs}
\usepackage{multirow}
\usepackage{stfloats}
\usepackage[hidelinks]{hyperref}
\def\BibTeX{{\rm B\kern-.05em{\sc i\kern-.025em b}\kern-.08em
    T\kern-.1667em\lower.7ex\hbox{E}\kern-.125emX}}
\begin{document}

\title{Economic Competition, EU Regulation, and Executive Orders: A Framework for Discussing AI Policy Implications in CS Courses}

\author{\IEEEauthorblockN{James Weichert}
\IEEEauthorblockA{
\textit{University of Washington}\\
Seattle, USA \\
jpw@cs.washington.edu}
\and
\IEEEauthorblockN{Hoda Eldardiry}
\IEEEauthorblockA{
\textit{Virginia Tech}\\
Blacksburg, USA \\
hdardiry@vt.edu}
}

\maketitle

\begin{abstract}
The growth and permeation of artificial intelligence (AI) technologies across society has drawn focus to the ways in which the responsible use of these technologies can be facilitated through AI governance. Increasingly, large companies and governments alike have begun to articulate and, in some cases, enforce governance preferences through AI policy. Yet existing literature documents an unwieldy heterogeneity in ethical principles for AI governance, while our own prior research finds that discussions of the implications of AI policy  are not yet present in the computer science (CS) curriculum. In this context, overlapping jurisdictions and even contradictory policy preferences across private companies, local, national, and multinational governments create a complex landscape for AI policy which, we argue, will require AI developers able adapt to an evolving regulatory environment. Preparing computing students for the new challenges of an AI-dominated technology industry is therefore a key priority for the CS curriculum.

In this discussion paper, we seek to articulate a framework for integrating discussions on the nascent AI policy landscape into computer science courses. We begin by summarizing recent AI policy efforts in the United States and European Union. Subsequently, we propose guiding questions to frame class discussions around AI policy in technical and non-technical (e.g., ethics) CS courses. Throughout, we emphasize the connection between normative policy demands and still-open technical challenges relating to their implementation and enforcement through code and governance structures. This paper therefore represents a valuable contribution towards bridging research and discussions across the areas of AI policy and CS education, underlining the need to prepare AI engineers to interact with and adapt to societal policy preferences. 
\end{abstract}


\section{Introduction}

\begin{figure}
    \centering
    
    \includegraphics[width=1\linewidth]{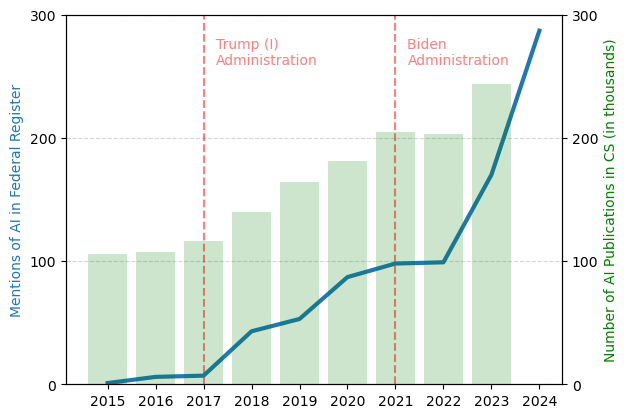}

    \caption{Mentions of ``artificial intelligence'' in the Federal Register, AI research publications in CS by year \cite{maslej_ai_2025}.}
    \label{fig:fed_reg}
\end{figure}

The intensifying focus of governments on the regulation of artificial intelligence (AI) technologies is evidenced both by major policy documents such as the EU AI Act \cite{eu_ai_act} or US executive orders \cite{exec_14110_2023, exec_14148_2025}, and in the steady increase in mentions of ``artificial intelligence'' in documents published in, for example, the US \emph{Federal Register} (Figure \ref{fig:fed_reg}). These developments, coupled with the abundance of available guidelines for `responsible AI' \cite{jobin_global_2019}, herald the emergence of \emph{AI policy} as an important field of study \cite{calo_artificial_2017, schiff_whats_2020, kim_toward_2024, weichert_perceptions_2025}. More concretely, we argue that policy and compliance considerations will have significant impacts on the AI development life cycle, necessitating AI developers able to navigate and actualize abstract policy demands (or reconcile conflicting ones). Yet, we find that discussions of AI policy developments are largely absent from the computer science (CS) curriculum \cite{weichert_i_2025, weichert_exploring_2025}, leaving critical competencies underdeveloped in the next generation of `AI practitioners' \cite{weichert_assessing_2025}.

Extending on our prior work exploring the development of AI policy curricular content for CS courses \cite{weichert_educating_2025, weichert_scoping_2025, weichert_ai_2025}, in this discussion paper we apply this background to (1) synthesize key AI regulatory developments in the United States and European Union (Section \ref{sec:landscape}); and (2) propose a novel framework to scaffold the integration of AI policy discussions in CS courses (Section \ref{sec:framework}). In doing so, we hope to offer a new (and practical) perspective to build on existing literature in the fields of AI policy, AI ethics, and CS education.

\section{The Emerging AI Policy Landscape}\label{sec:landscape}

In order to construct the framework we describe in Section \ref{sec:framework}, we aim to first synthesize similarities and key themes characterizing the development and enactment of AI policy in the United States and the European Union (EU). This section therefore serves two purposes: first, summaries of the policy landscapes across these two jurisdictions provide helpful context for instructors wishing to discuss current AI policy developments in computing courses; and second, an understanding of these policy contexts is necessary to anticipate the ways in which AI developers might interact with a growing AI regulatory ecosystem.

We focus on the US and EU because they represent (along with the People's Republic of China), the most important state actors in the AI domain \cite{roberts_achieving_2021, weichert_educating_2025}. Moreover, we believe these two polities will have the most immediate influence on the work of AI developers in the US, and as such they are the most relevant to us as US-based AI educators. Nevertheless, we emphasize that what actors are `relevant' will ultimately vary on geopolitical context, and suspect that encouraging students to contrast approaches to AI policy across countries (e.g., between the US and their home country) will only enhance the depth of classroom discussion.    

\subsection{United States}

Existing literature \cite{roberts_achieving_2021, parinandi_investigating_2024, weichert_perceptions_2025} emphasizes the heterogeneity of AI policy efforts in the US, and the infancy of enforceable regulation. Whereas the European Union made steady progress towards the creation of the \emph{EU AI Act} from 2016 through the law's enactment in 2024 \cite{delvaux_draft_2016, madiega_artificial_2024}, the US has yet to enact comprehensive federal legislation on AI. Instead, US efforts in this area have tended to comprise non-binding policy statements such as the National Institute of Standards and Technology (NIST) \emph{AI Risk Management Framework} \cite{national_institute_of_standards_and_technology_artificial_2023}, the \emph{American AI Initiative} from President Trump's first term \cite{noauthor_national_2021, roberts_achieving_2021}, or the Biden White House's \emph{Blueprint for an AI Bill of Rights} \cite{office_of_science_and_technology_policy_blueprint_2022}. Furthermore, as Parinandi et al.~\cite{parinandi_investigating_2024} show, AI policy proposals on the state level exhibit ideological and partisan divides, with Democratic lawmakers voting more favorably towards consumer protection-oriented AI legislation compared to Republican lawmakers, who tend emphasize promoting economic growth and competitiveness.

These fault lines have recently become clear on the federal level as well through opposing executive orders. In 2023, then-President Biden signed Executive Order (EO) 14110 on \emph{Safe, Secure, and Trustworthy Development and Use of Artificial Intelligence} \cite{exec_14110_2023}, the most comprehensive and actionable statement on federal AI policy to date \cite{coglianese_people_2023}. EO 14110 established eight policy areas—including \emph{safety and security}, \emph{consumer protection}, and \emph{privacy}—and described the Biden administration's AI policy across nine paragraphs. Yet this executive order was rescinded on January 20, 2025 by President Trump \cite{exec_14148_2025}, underscoring the recent ideological polarization over AI. As as replacement, Trump signed EO 14179, titled \emph{Removing Barriers to American Leadership in Artificial Intelligence}, which established new US AI policy in a single sentence: ``It is the policy of the United States to sustain and enhance America's global AI dominance in order to promote human flourishing, economic competitiveness, and national security,'' \cite{exec_14179_2025}. The focus on economic and national security motivations—over considerations for safety and human rights—is conspicuous, as is the assertion that the US can claim ``global AI dominance'' 
(especially in light of the challenge posed by the newly-released \textit{Deepseek} LLM from China to American AI chatbot hegemony \cite{mhlanga_ai_2025}). The order also directs agencies to identify ``actions taken pursuant to Executive Order 14110 that are or may be inconsistent with'' this new single-sentence policy statement. In other words, the new administration sees a focus on ``safe, secure, and trustworthy'' AI as a ``barrier`` to continued American economic success in the AI sector. It is this newest formulation of policy which will most heavily influence any legislative developments for the remainder of the Republican-controlled 119th US Congress.

\subsection{European Union}

AI policy development in the European Union has concentrated in Brussels, with the European Commission and its High-Level Expert Group (HLEG) taking the lead in the drafting of the \emph{EU AI Act} \cite{eu_ai_act}, which received final assent from the European Parliament in March 2024 \cite{madiega_artificial_2024}. The Act defines an ``AI system'' as one that ``is designed to operate with varying levels of autonomy and that may exhibit adaptiveness after deployment'' \cite{eu_ai_act}, thereby encompassing a broad range of algorithms. These systems are organized by risk level, with \emph{high-risk} practices related to product safety (Chapter III) subject to additional risk management, transparency, and oversight requirements. AI practices deemed `unacceptable risks'—including social scoring, subliminal manipulation, and `real-time' biometric surveillance—are prohibited outright (with some exceptions for law enforcement) \cite{veale_demystifying_2021}. Moreover, providers of `general-purpose' AI models, meaning those which are ``capable of competently performing a wide range of distinct tasks,'' are subject to additional transparency requirements, including the disclosure of information relating to the model's architecture and training data and the institution of a ``policy to comply with Union law on copyright,'' \cite{eu_ai_act}. Recitals 98 and 99 specifically incorporate models with ``at least a billion of parameters'' and ''large generative'' models into the `general-purpose' definition \cite{eu_ai_act}, casting a wide net.

\subsection{Themes}

Below, we summarize three implications of the developing AI policy landscape which we synthesize from our review of government AI regulatory efforts in the US and EU. We view these themes as relevant to a variety of jurisdictions, contexts, and roles. More to the point, these implications underscore the importance of integrating discussions about AI governance in `technical' AI curricula. As these (often vague) government policies emphasize, the effective technical regulation and oversight of AI requires closer coordination between policy-setter and engineer. 

\paragraph{The Rapidly Changing AI Landscape}

The recent about-face in US AI policy evidenced by President Trump's rescission of President Biden's EO 14110 makes clear the unusual volatility of this policy area, at least in this early stage of US policymaking on AI. In the absence of a bipartisan congressional consensus on concrete AI regulation, we should expect continued oscillation between opposing focal points of AI policy. Research \cite{hine_artificial_2022, parinandi_investigating_2024} and observation \cite{exec_14110_2023, exec_14179_2025} both suggest that, whereas Democratic policymakers will place greater emphasis on the cultivation of ``safe'' and ``trustworthy'' AI, their Republican counterparts will focus almost exclusively on promoting US economic and national security ``AI dominance.'' The swings in US policy to be expected in the foreseeable future will require AI companies and practitioners to quickly adapt to shifting regulatory demands. We can therefore conclude that AI developers possessing a degree of familiarity with this AI policy landscape will be better equipped to manage these oscillations, and therefore more desirable to AI companies.

\paragraph{Technical Compliance with Non-Technical Specifications} A primary concern with respect to the effective governance of AI is how to translate abstract policy goals to concrete controls on AI models. For example, how should we define and detect ``algorithmic discrimination'' (EO 14110) in the context of AI decision-making? Likewise, what should be done to ensure generative AI models ``comply'' with copyright law (EU AI Act) when the process for \emph{how} models produce generative outputs is not fully understood? These challenges will require AI developers to possess both technical AI development skills and an AI policy `vocabulary' to understand how vague policy desires can and should be implemented technically. Similarly, there are some areas of the AI regulatory frontier that do not (yet) have technical solutions. For instance, a requirement to identify and label AI-generated context would require an AI watermark that cannot be circumvented, which does not currently exist. In these cases, developers must recognize where the `technical frontier' lies, and engage in conversations with policymakers to set realistic and achievable expectations. Questions of accuracies, tolerances, and average outcomes will be of critical importance here.

\paragraph{Cross-Border Implications}

Even for engineers working in a US market, the influences of external regulatory efforts cannot be ignored. The technology economy is global, so economically-viable technologies can no longer be developed in the vacuum of a single country's regulatory context. The EU's General Data Protection Regulation (GDPR) provides the best example of the wide-reaching impact of recent EU policy. The cost of ensuring GDPR compliance within EU jurisdiction incentivizes companies to extend this compliance to its users across all jurisdictions \cite{gstrein_extraterritorial_2021}, resulting in the effective application of the GDPR in non-EU countries, evidenced by the GDPR cookie pop-ups shown to US users on (some) US company websites. This is an example of what Anu Bradford \cite{bradford_brussels_2012} terms the ``Brussels Effect,'' whereby the EU ``sets the global rules across a
range of areas.'' The case study of the GDPR's reach beyond EU borders makes us believe that the timeliness of the EU AI Act will have similar effects. We wonder if American AI companies in particular might begin to adopt AI Act compliance standards even for the US market in the absence of an AI regulatory consensus in Washington.

\begin{table*}[htb]

    \centering
    \caption{A framework for evaluating the implications of AI policy.}
    \label{tab:questions}
    
    \begin{tabular}{l|p{4.5cm}|p{4.5cm}|p{4.5cm}}
        \toprule
         & \multicolumn{1}{c|}{\textbf{Socio-Political}} & \multicolumn{1}{c|}{\textbf{Translational}} & \multicolumn{1}{c}{\textbf{Technical}} \\
         \midrule
         \multirow{7}{*}{\textbf{\emph{SCOPE}}} & \begin{itemize}
            \item What ethical principles is this policy intended to reinforce?
             \item To which political (or social) jurisdictions does this policy apply?
             \item What incentives or penalties does this policy carry?
         \end{itemize} & \begin{itemize}
             \item What technical specifications do these ethical principles require? 
             \item How do these jurisdictions translate to technical user groups?
             \item How do these incentives/penalties apply to the digital space? 
         \end{itemize} & \begin{itemize}
             \item What algorithms/systems satisfy these technical specifications?
             \item How should user permissions be configured? 
             \item What mechanisms/protocols can be used to enforce the policy?
         \end{itemize} \\
         \midrule
         
         \multirow{7}{*}{\textbf{\emph{POWER}}} & \begin{itemize}
             \item Which individuals and/or organizations created this policy?
             \item Who benefits from this policy? Is anyone (disproportionately) harmed by this policy?
         \end{itemize} & \begin{itemize}
             \item Does this policy carry any implied values or assumptions?   
             \item How do the demographics and experiences of the developers compare to the user population? Are there perspectives missing?
         \end{itemize} & \begin{itemize}
             \item Who will be responsible for overseeing the implementation of this policy?
             \item What technical safeguards should be implemented to prevent harm to users?
         \end{itemize} \\
         \midrule
         
         \multirow{8}{*}{\textbf{\emph{AGENCY}}} & \begin{itemize}
             \item Who was consulted (explicitly or implicitly) in shaping this policy? Who was left out?
             \item What opportunities exist to influence this policy?
         \end{itemize} & \begin{itemize}
             \item What shape does (or should) the interaction between policymakers and developers regarding the implementation of this policy take?
             \item What role can (or should) developers play in shaping the implementation of this policy?
         \end{itemize} & \begin{itemize}
             \item  Are there technical considerations or limitations that conflict with policy expectations?
             \item How can the faithful technical implementation of the policy be measured?
         \end{itemize} \\
         \bottomrule
    \end{tabular}
\end{table*}

\section{Framework}\label{sec:framework}

Based on this synthesis of the developing AI policy landscape—and building on our previous research highlighting gaps in the `AI curriculum` \cite{weichert_i_2025, weichert_exploring_2025, weichert_scoping_2025} and developing AI policy curricular content for CS courses \cite{weichert_educating_2025, weichert_ai_2025}—in this section we propose a framework to organize and facilitate the integration of discussions on AI policy in computing courses. 

\subsection{Perspectives}

\begin{figure}
    \centering
    \includegraphics[width=1\linewidth]{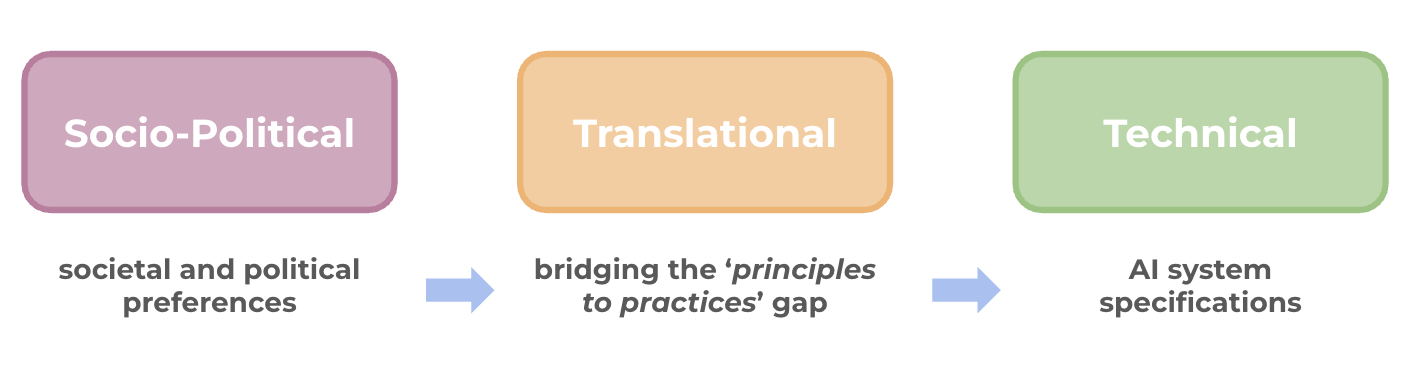}
    \caption{A continuum of perspectives for AI policy analysis.}
    \label{fig:perspectives}
\end{figure}

We decompose discussions about the implications of AI policies across a continuum of perspectives ranging from \textbf{socio-political} considerations to \textbf{technical} ones (Figure \ref{fig:perspectives}). Whereas the former encompasses normative preferences shaped by the representation of society through democratic institutions, the latter is concerned with the implementation of technical specifications in AI systems. As we identify in Section \ref{sec:landscape} and in our prior work \cite{weichert_scoping_2025}, aligning policy preferences with technical specifications is rarely straightforward. In some cases, technical limitations may necessitate policy compromises for expediency.

To bridge the gap, we propose a \textbf{translational} `perspective' focused on how to transition from `principles to practices' \cite{schiff_principles_2020, kim_toward_2024}. This transition requires policymakers and AI developers to collaborate in formulating technical specifications to match the letter (and spirit) of policy through the standardization of definitions; the quantification of desired accuracies, tolerances, and levels of reliability; and the consideration of `edge cases'. The guiding questions in Section \ref{sec:guiding-questions} thus aim to encourage students to consider normative policy demands in the context of the technical intricacies of state-of-the-art AI systems.

\subsection{Dimensions}

Orthogonal to the three perspectives described above, three dimensions—\emph{scope}, \emph{power}, \emph{agency}—act as lenses through which to segment concerns about AI policy implications:

\paragraph{Scope} The scope (and content) of an AI policy is perhaps the most straightforward to analyze—what does the policy require or prohibit, and of whom? Yet even here there are intricacies revealed by changing perspectives. A requirement for `maintaining user privacy' requires establishing working definitions for `privacy' (e.g., what information should not be revealed or inferable and by whom?) and identifying algorithms and protocols (e.g., differential privacy) for implementing this level of guarantee in the software. In interrogating the finer intricacies of policy requirements, students are coaxed to connect ethical principles (e.g., privacy) with particular technical mechanisms for achieving these principles. The current limitations of these mechanisms define what we term the `technical frontier' of responsible AI \cite{weichert_scoping_2025}, reinforcing the still-open research questions at the heart of abstract ethical preferences.

\paragraph{Power} Building on previous work identifying the analysis of power structures as important yet underdeveloped competencies for the `AI curriculum' \cite{weichert_scoping_2025}, we emphasize \emph{power} as a central focus of discussions around AI policy. Students should be encouraged to identify the actors (individuals or institutions) involved with the creation and implementation of AI policy, examine their motivations, and consider groups and perspectives that are underrepresented in this process.  

\paragraph{Agency} We are inspired to include a focus on agency by the observation of Padiyath \cite{padiyath_realist_2024} that, ``When engaging with ethical dilemmas set in the workplace, students are often worried about the lack of agency they have to navigate these dilemmas.'' Addressing the roles and influence of AI practitioners at all levels—from junior AI engineers to chief technology officers—in shaping the responsible development and use of AI technologies is therefore paramount. The \emph{agency} dimension encourages students to imagine avenues for influencing AI policy at various scales, from engineering team meetings to participation in democratic political processes.

\subsection{Guiding Questions}\label{sec:guiding-questions} 

Our framework is most comprehensively embodied through the grid of guiding questions shown in Table \ref{tab:questions}. Each cell contains discussion questions relating to the intersection of a particular dimension with a particular perspective. For example, an AI policy may only specify the social or political groups to which it applies (\textbf{socio-political} perspective on the policy's \emph{scope}), but the \textbf{translational} step requires mapping these groups to rigid sets of users. A \textbf{technical} challenge then arises in ensuring that the AI system can accurately and reliably identify users belonging to the groups described by the policy (e.g., minors).

\section{Conclusion}

The framework we outline above provides a robust starting point for classroom discussions and assignments involving the evaluation or implementation of AI policy. At the same time, this structure is deliberately flexible and context-agnostic, easily adapting to different curricula (e.g., `Natural Language Processing' vs. `CS Ethics Seminar') and policy environments. Concretely, we see the guiding questions we propose as helpful for encouraging students to consider other (underrepresented) perspectives, evaluate their own agency to influence the responsible use of AI, and question how abstract policy demands can be translated to specific technical interventions. This discussion paper is only the first iteration of this framing, condensing multiple strands of our prior research into a compact format, so we welcome feedback and extensions on this work.

\subsection{Future Work}

Two areas stand out in particular as promising opportunities for future research. First, our framework could be coupled with specific curricular activities to target particular intersections of perspective and dimension. For example, what exercises could be employed to facilitate the inclusion of underrepresented perspectives in the design of AI systems (i.e., a \textbf{translational} perspective on \emph{power})? Second, the translation of abstract policy preferences to concrete technical specifications calls to mind the process of requirements engineering. The application of research and practice from this field to the AI `principles to practices' gap therefore appears both novel and valuable.

\bibliographystyle{IEEEtran}
\bibliography{references}

\end{document}